\documentclass[floatfix,aps,amsmath,nofootinbib,onecolumn,11pt]{revtex4-1}
\def \be {\begin{equation}} 
\def \ee {\end{equation}} 
\def \bea {\begin{eqnarray}} 
\def \eea {\end{eqnarray}} 

\usepackage{graphicx}
\usepackage{dcolumn}
\usepackage{bm}
\usepackage{epsfig} 
\usepackage{amsfonts}
\usepackage{amsmath}
\usepackage{amssymb}
\usepackage[usenames]{color}
\usepackage[dvipsnames]{xcolor}
\usepackage[unicode, colorlinks=true, linkcolor=linkcolor, citecolor=linkcolor, filecolor=linkcolor,urlcolor=linkcolor, pdfusetitle]{hyperref}

\hypersetup{colorlinks,citecolor=blue,linkcolor=blue,urlcolor=blue}
\hypersetup{final=true}

\begin{document}

\title{Testing the isotropy of cosmic acceleration with Pantheon+ and SH0ES datasets: A cosmographic analysis}

\author{Carlos A. P. Bengaly}
\email{carlosbengaly@on.br}
\affiliation{Observat\'orio Nacional, 20921-400, Rio de Janeiro - RJ, Brazil}

\author{C\'assio Pigozzo}
\email{cpigozzo@ufba.br}
\affiliation{Instituto de F\'isica, Universidade Federal da Bahia, 40210-340, Salvador - BA, Brazil}

\author{Jailson S. Alcaniz}
\email{alcaniz@on.br}
\affiliation{Observat\'orio Nacional, 20921-400, Rio de Janeiro - RJ, Brazil}

\date{\today}

\begin{abstract}

We use a recent Pantheon+SH0ES compilation of Type Ia Supernova distance measurements at low-redshift, i.e., $0.01 \leq z \leq 0.10$, in order to investigate the directional dependency of the deceleration parameter ($q_0$) in different patches ($60^{\circ}$ size) across the sky, as a probe of the statistical isotropy of the Universe. We adopt a cosmographic approach to compute the cosmological distances, fixing $H_0$ and $M_B$ to reference values provided by the collaboration. By looking at 500 different patches randomly taken across the sky, we find a maximum $\sim 3\sigma$ CL anisotropy level for $q_0$, whose direction points orthogonally to the CMB dipole axis, i.e., $(RA^{\rm SN},DEC^{\rm SN}) = (267^{\circ},6^{\circ})$ vs $(RA^{\rm CMB},DEC^{\rm CMB}) = (167^{\circ},-7^{\circ})$. We assessed the statistical significance of those results, finding that such a signal is expected due to the limitations of the observational sample. These results support that there is no significant evidence for a departure from the cosmic isotropy assumption, one of the pillars of the standard cosmological model. 

\end{abstract}


\maketitle

\section{Introduction}\label{sec:intro}

One of the foundations of the current standard cosmological model,  also known as  $\Lambda$ - Cold Dark Matter ($\Lambda$CDM) scenario, is the assumption of the validity of the Cosmological Principle (CP) at large scales~\cite{Goodman:1995dt,Clarkson:2010uz,Maartens:2011yx,Clarkson:2012bg,Aluri:2022hzs}. In other words, it states that the Universe should appear statistically homogeneous and isotropic at those scales, so that we can measure cosmological distances and ages using the Friedmann-Lema\^itre-Robertson-Walker (FLRW) metric. Recent analysis of redshift surveys of galaxies and quasars indicate that there is indeed a transition scale from a locally inhomogeneous Universe to a smoother, statistically homogeneous one, as described by the CP, at about 70-120 Mpc~\cite{Laurent:2016eqo,Ntelis:2017nrj,Goncalves:2017dzs,Goncalves:2020erb,Kim:2021osl,Andrade:2022imy}. 

Hence, we must test the assumption of the cosmological isotropy as well in order to confirm (or rule out) the CP as a valid physical assumption. If ruled out, the standard model would require a profound reformulation of its basic hypotheses, including the physical origin of the mechanism behind cosmic acceleration. One possible way to test such a hypothesis involves analyzing the cosmological parameters' directional dependency, as estimated from the luminosity distance measurements of Type Ia Supernova (SNe). This method has been explored since the release of the earliest SN compilations, using a variety of approaches, yielding inconclusive results due to the limited sampling of objects -- in terms of small number of SNe available, distance measurement uncertainties, and especially uneven sky coverage~\cite{Kolatt:2000yg,Schwarz:2007wf,Antoniou:2010gw,Cai:2011xs,Bahr-Kalus:2012yjc,Zhao:2013yaa,BeltranJimenez:2014otq,Chang:2014wpa,Bengaly:2015dza,Javanmardi:2015sfa,Deng:2018jrp,Andrade:2018eta,Colin:2019opb,Zhao:2019azy,Kazantzidis:2020tko,Hu:2020mzd,Salehi:2020hek,Zhao:2021fcp,Krishnan:2021jmh,Horstmann:2021jjg,Rahman:2021mti,Dhawan:2022lze}. The latest SN compilation, namely the Pantheon+SH0ES data-set~\cite{Brout:2022vxf} (see also~\cite{Scolnic:2021amr,Riess:2021jrx}), provides 1701 light curve measurements of 1550 SN objects. This corresponds to a significant  improvement from previous figures of earlier SN compilations -- for instance, the previous Pantheon and JLA compilations comprised 1048 and 740 SN distance measurements, respectively. However, recent analyses still yield inconclusive results regarding the validity of cosmic isotropy, depending on the sample selection and the methodology adopted~\cite{Sorrenti:2022zat,Cowell:2022ehf,Pasten:2023rpc,McConville:2023xav,Perivolaropoulos:2023tdt,Tang:2023kzs,Hu:2023eyf}. 

Given this scenario, we look further at the isotropy of the Pantheon+SH0ES data. In this case, we carry out a directional analysis of the deceleration parameter ($q_0$) across the celestial sphere. Similar approaches were adopted in~\cite{McConville:2023xav,Hu:2023eyf}, although the authors focused on the $H_0$ and $\Omega_{\rm m}$ parameters, as given by the standard model, i.e., within the flat $\Lambda$CDM framework. Instead, we rely on a cosmographic description of the Universe so that no further assumptions on its dynamic content, e.g., the nature of dark matter and dark energy, are needed -- as long as we restrict our analysis to the lower redshift threshold of the sample. We also estimate the statistical significance of our results in light of the assumptions made on data analysis, or the non-uniformity of the SN sky distribution. 

The paper is structured as follows: Section II is dedicated to explaining our method and the data selection and preparation. Section III presents the results obtained from this
method, along with the statistical significance tests. Section IV provides the discussion and our concluding remarks.

\section{Method}

\begin{figure*}[!ht]
\centering
\includegraphics[width=0.82\linewidth, height=7.5cm]
{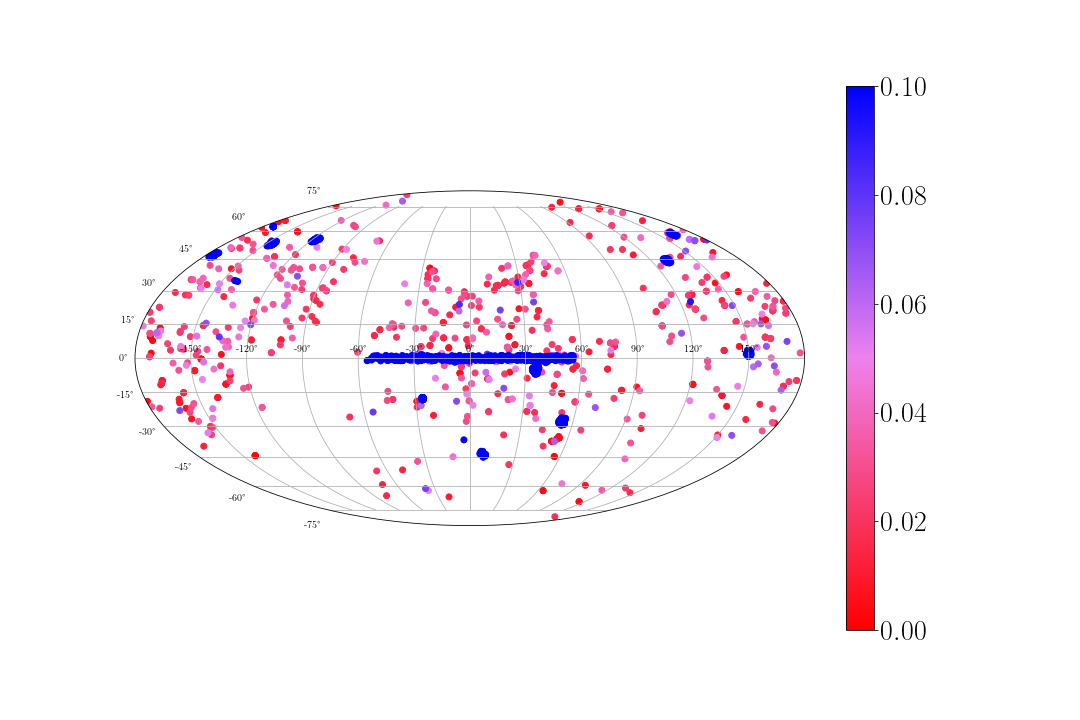}
\includegraphics[width=0.82\linewidth, height=7.5cm]
{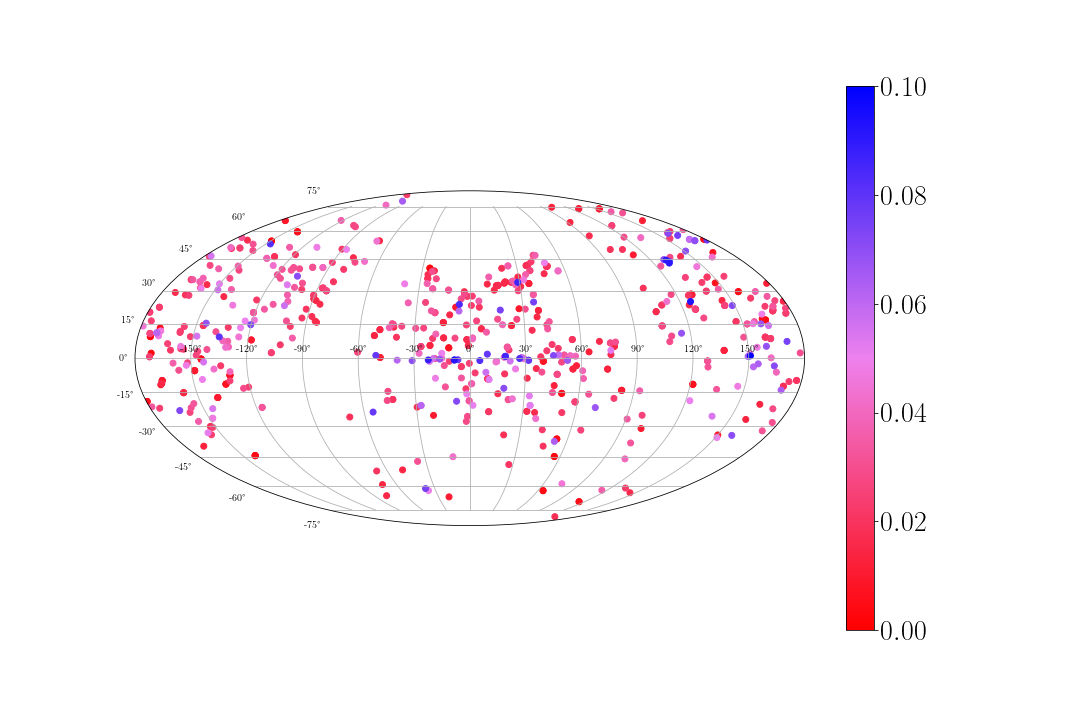}
\caption{A Mollweide projection of the SN positions in the sky. In the {\it upper panel}, we assign their positions in such a way that the colour bar is truncated in $z=0.1$, so that all SNe with $z>0.1$ are shown in blue. On the other hand, the {\it lower panel} displays the SNe that made our imposed redshift cut, i.e., $0.01<z<0.10$, in addition to the SNe in Cepheid host galaxies, for a colour range $z=0$ (reddest) to $z=0.1$ (bluest). Both maps are projected in celestial sky coordinates $(RA,DEC)$.}
\label{fig:SN_mollview_projections}
\end{figure*}

\subsection{Data Preparation}

We use the Pantheon+SH0ES compilation, as retrieved from the {\sc Github} repository\footnote{\url{https://github.com/PantheonPlusSH0ES/DataRelease}}. This data set consists on 1701 light curve measurements of 1550 distinct SNe within the $0.001 < z < 2.26$ redshift range, which also comprises 77 data points from Cepheid host galaxies at very low redshifts, i.e., $0.00122 < z < 0.01682$. Those measurements are designated in the data release as a boolean variable, so that we use the Cepheid host distances provided in those cases, instead of the distance modulus measured by the corresponding SN. Note we use the redshift given in the {\sc zHD} variable, since it includes the correction from the heliocentric to the Cosmic Microwave Background rest-frame, as well as further corrections due to the nearby SN peculiar velocities~\cite{Peterson:2021hel,Carr:2021lcj}. Still, we caution that the effect of those peculiar velocities can lead to spurious anisotropy in the Hubble Flow, if not properly accounted for~\cite{Hellwing:2016pdl,Bengaly:2018uqp,Kalbouneh:2022tfw,Maartens:2023tib,Kalbouneh:2024szq}. Although some doubts have been cast on the modelling of the SN peculiar velocities in this sample (see~\cite{Sorrenti:2022zat}), we will not attempt to reevaluate their effects on the isotropy test we perform in this work.

It is important to avoid possible biases from the original sample as much as possible,
because it consists of an SN compilation from observations performed by various surveys
that have (or had) different designs and sky coverage. Thus, the original SN compilation is expected to be affected mainly by uneven sky coverage. Due to this, we impose two different redshift cutoffs before we proceed to the isotropy test. The first one consists of an upper redshift cutoff at the $z > 0.10$ range, since the SN sky distribution becomes much more sparse at higher redshifts (see upper panel of Fig.~\ref{fig:SN_mollview_projections} for a visual explanation). This is something expected, since those objects were observed by spectroscopic surveys designed to cover only those specific regions of the sky, in order to obtain the highest-resolution spectroscopy possible. Also, we can safely use cosmographic expansion on equation \ref{eq:DL} since it is a good approximation in that redshift range. Secondly, we impose a lower redshift cutoff at $z < 0.01$, except for the distance measurements within Cepheid host galaxies. Hence, we end up with a working sample of 697 SN data points, as displayed in the lower panel of Fig.~\ref{fig:SN_mollview_projections}.

\subsection{Estimator}

We assume a cosmographic expression of the luminosity distance~\cite{Weinberg:1972kfs,Visser:2003vq,Cattoen:2007id}
\begin{eqnarray}
\label{eq:DL}
D_{\rm L}(z) = (c/H_0)\left[z + (1-q_0)z^2/2\right], 
\end{eqnarray}
where $H_0$ and $q_0$ stand for the Hubble Constant and the deceleration parameter, respectively, and $D_{\rm L}(z)$ is given in Mpc. As the distance modulus definition reads
\begin{eqnarray}
\label{eq:mu}
\mu_{\rm model}(z) \equiv m - M = 5\log_{10}{(D_{\rm L}(z)/\mathrm{Mpc})} + 25,
\end{eqnarray}
we obtain the best-fit for the $q_0$ through a $\chi^2$ minimization, as given by
\begin{eqnarray}
\label{eq:chi2}
\chi^2 = \vec{\delta^{\rm T}} (C_{\rm stat+sys})^{-1} \vec{\delta},
\end{eqnarray}
so that $C_{\rm stat+sys}$ denotes the full SN covariance matrix, and
\begin{eqnarray}
\label{eq:deltacases}
\delta_{\rm i} = 
\begin{cases}
m_{\rm i} - M - \mu_{\rm i}, \quad i \in {\rm Cepheid} \, \rm{Hosts} \\
m_{\rm i} - M - \mu_{\rm model}(z_i) \quad {\rm otherwise} \,.
\end{cases}
\end{eqnarray}

We adopt the following estimator to test the isotropy of the local cosmic acceleration by a similar fashion of~\cite{McConville:2023xav,Hu:2023eyf}, as follows
\begin{eqnarray}
\label{eq:deltaq0}
\Delta_{q0} = \frac{|{q^{\rm (``north cap")}_0}|-|{q^{\rm (``south cap")}_0}|}{\sigma_{q^{\rm (``north cap")}_0}^2 + \sigma_{q^{\rm (``south cap")}_0}^2},
\end{eqnarray}
where $\Delta_{q0}$ is given in units of confidence level (CL). We compute the best-fitted value for the deceleration parameters at opposite patches across the entire celestial sphere ($60^{\circ}$ size) along a specific axis randomly selected across the sky. A total of 500 axes were taken in our analysis. In Eq.~\ref{eq:deltaq0}, these quantities are denoted by $q^{\rm (``north cap")}_0$, and $q^{\rm (``south cap")}_0$, correspondingly, while $\sigma_{q^{\rm (``north cap")}_0}^2$ and $\sigma_{q^{\rm (``south cap")}_0}^2$ provides their respective uncertainties at $1\sigma$ CL. Explicitly, we do so by computing likelihoods of the $q^{\rm (``north cap")}_0$, and $q^{\rm (``south cap")}_0$ according to $\mathcal{L} \propto \exp{-\chi^2/2}$, where $\chi^2$ corresponds to Eq.~\eqref{eq:chi2}. Then, we use the {\sc curvefit} routine of the {\sc SciPy} module to fit a Gaussian curve to the likelihoods at each cap, and we take its mean value and standard deviation as our corresponding $q_0$ and $\sigma_{q_0}$ in each ``north/south" cap, respectively, in Eq. ~\eqref{eq:deltaq0}. 

Finally, we stress two more points: (i) We perform our analysis within $60^{\circ}$ size spherical caps, instead of the typical choice of $90^{\circ}$ size ones -- which encompasses an entire hemisphere -- for the sake of computational time. Nonetheless, we expect that this choice should not impact our conclusions, as the results obtained from $60^{\circ}$ caps show good agreement with those assuming $90^{\circ}$ caps -- see Fig. 18 in~\cite{Hu:2023eyf}; (ii) The fact that we are assuming a cosmographic expansion, rather than the exact expression for the luminosity distance in the $\Lambda$CDM framework, does not bias our inferences. The relative difference between the distance modulus predicted by both cases, assuming self-consistent cosmological parameters, i.e., $q_0=-0.499$ for former, and $\Omega_{\rm m}=0.334$ for the latter -- and both assuming $H_0 = 73.3 \, \mathrm{km} \, \mathrm{s}^{-1} \, \mathrm{Mpc}^{-1}$ -- only differ at about $0.35$\% at $z = 0.1$, which is the highest redshift covered by our SN sub-sample. Since the uncertainties of the SN measurements are typically larger than those values, our analysis and conclusions should not be affected by that.

\section{Results}

\begin{figure*}[!ht]
\centering
\includegraphics[width=0.49\linewidth, height=6.3cm]{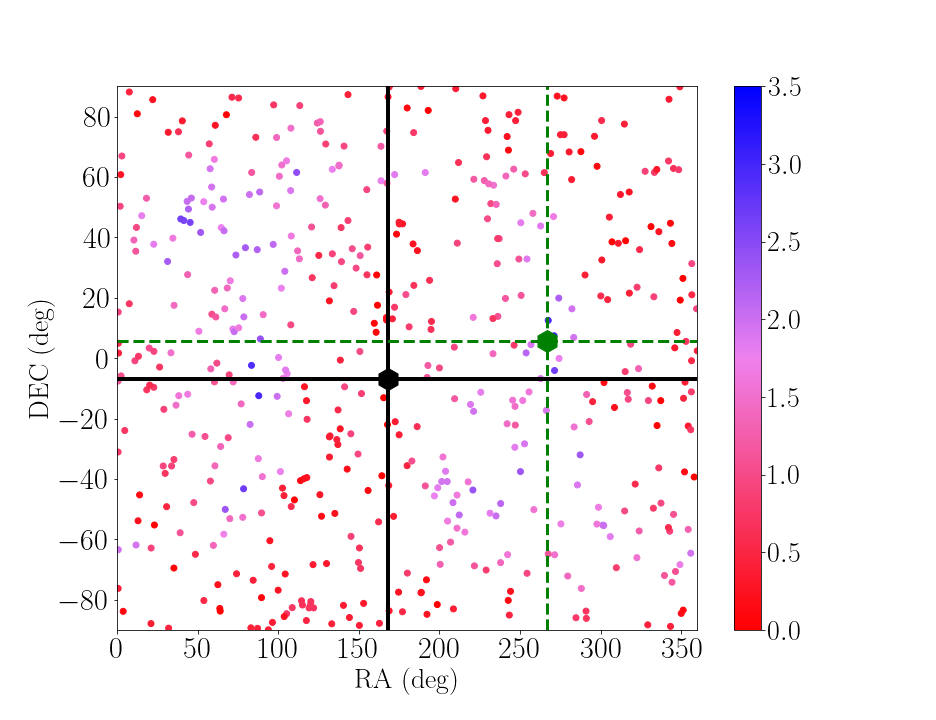} 
\includegraphics[width=0.49\linewidth, height=6.3cm]{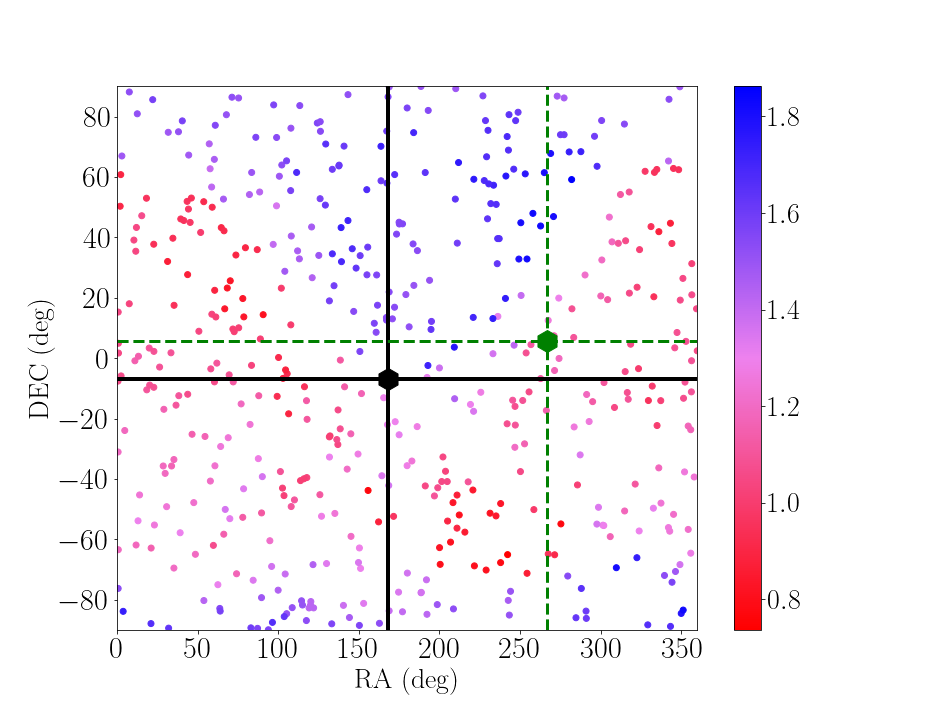} 
\caption{{\it Left panel}: A Cartesian projection map of the $\Delta_{q0}$ analysis. We take 500 random directions on our Pantheon+SH0ES selected sub-sample, as represented by each coloured point in this map, for a colour bar at the range $0 < \Delta_{q0} < 3.5$ (C.L.). The green diamond mark displays the maximum anisotropy direction, found along the $(RA^{\rm SN},DEC^{\rm SN}) = (267^{\circ},6^{\circ})$ axis, whereas the black dotted mark represents the axis of the CMB dipole direction, i.e., $(RA^{\rm CMB},DEC^{\rm CMB}) = (167^{\circ},-7^{\circ})$. {\it Right panel}: Same as left panel, but rather for the reduced $\chi^2$ value, $\chi^2_{\rm red}$, obtained in each of those random sky directions.}
\label{fig:deltaq0_map}
\end{figure*}

\begin{figure*}[!ht]
\centering
\includegraphics[width=0.48\linewidth, height=6.5cm]{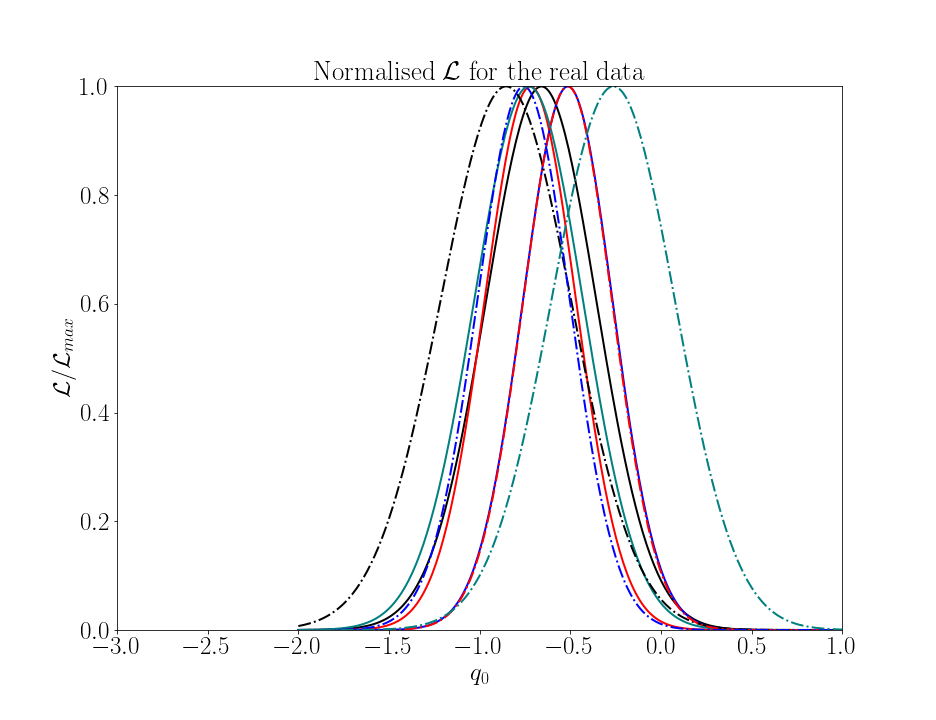}
\includegraphics[width=0.48\linewidth, height=6.5cm]{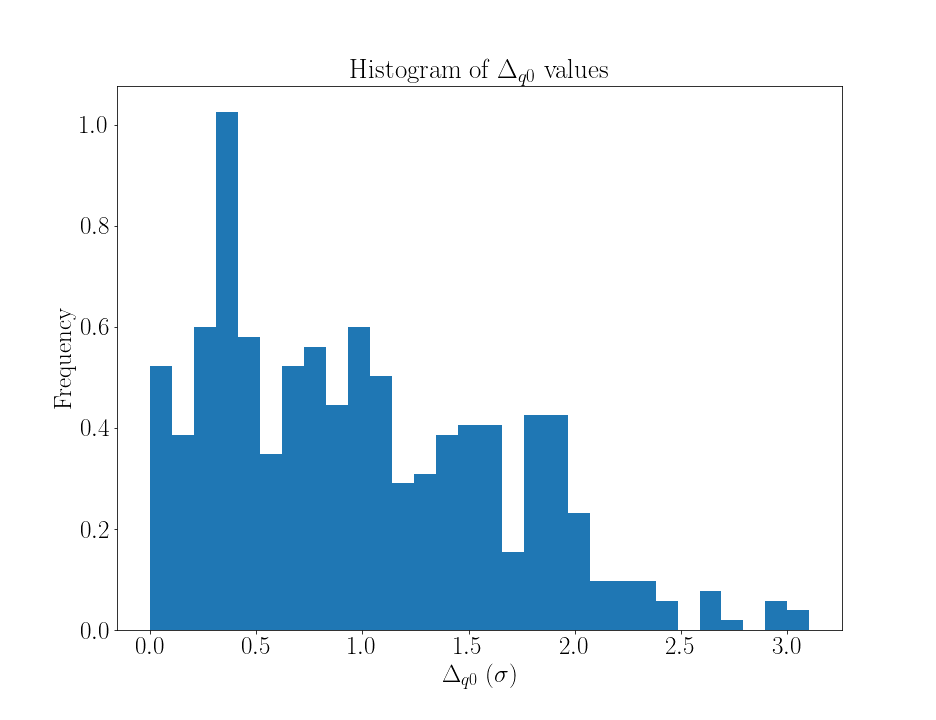}
\caption{{\it Left panel}: Examples of the $q_0$ normalised likelihood from 4 different directions across the sky. The black, red, blue and teal solid (dashed-dotted) curves represent the ``northern" (``southern") counterpart of each of those directions. {\it Right panel}: A histogram of $\Delta_{q0}$ values, given in function of C.L. ($\sigma$).}
\label{fig:q0lik_deltaq0}
\end{figure*}

\subsection{Real data analysis}

We show the result of the $\Delta_{q0}$ analysis in the left panel of Fig.~\ref{fig:deltaq0_map}. We represent it as a map in Cartesian coordinates of the $\Delta_{q0}$ values obtained across the 500 randomly selected directions in the sky for the Pantheon+SH0ES SN sample. Here, $\Delta_{q0}$ is given in units of standard deviations. The green diamond mark displays the maximum anisotropy direction found, along the $(RA^{\rm SN},DEC^{\rm SN}) = (267^{\circ},6^{\circ})$ axis, as represented by the green dashed lines, while the black solid lines and black dotted mark represents the CMB dipole direction, i.e., $(RA^{\rm CMB},DEC^{\rm CMB}) = (167^{\circ},-7^{\circ})$. We report a maximum $\Delta_{q0}$ value of $\Delta_{q0}=3.06$ in this case.

For the sake of consistency, we also display the reduced $\chi^2$ value in the right panel of the same figure, as defined by $\chi^2_{\rm red} \equiv \chi^2/{\rm dof}$, where $\mathrm{dof}$ denotes the number of degrees of freedom in each celestial cap, and $\chi^2$ is given by Eq.~\ref{eq:chi2}. We can see that the bluest dots, which denote the poorest $q_0$ fits to the data, does not coincide with the bluest regions in the $\Delta_{q0}$ map. This result indicates that the largest $\Delta_{q0}$ values do not occur due to a poor fit of the data, and that the largest $\chi^2_{\rm red}$ values are rather due to under-sampling of SN objects in the corresponding sky regions. 

Accordingly, in the left panel of Fig.~\ref{fig:q0lik_deltaq0}, we present examples of the normalised $q_0$ likelihood functions obtained from four different directions in the sky to visualize better the fits. In this case, different colours denote different sky directions, whereas the solid and dashed-dotted curves stand for the ``northern" and ``southern" counterpart of each direction respectively. In addition, the right panel of the same figure shows a histogram of the $\Delta_{q0}$ values, as given in confidence levels. We find that this distribution peaks at $ \Delta_{q0} \sim 0.5$, which corresponds to the most frequent value for the $q_0$ anisotropy level, and that its maximum value is $\Delta_{q0}=3.06$ -- as also shown in the left panel of Fig.~\ref{fig:deltaq0_map}. 
 
\subsection{Statistical significance of the results}

\begin{figure*}[t]
\centering
\includegraphics[width=0.48\linewidth, height=6.0cm]{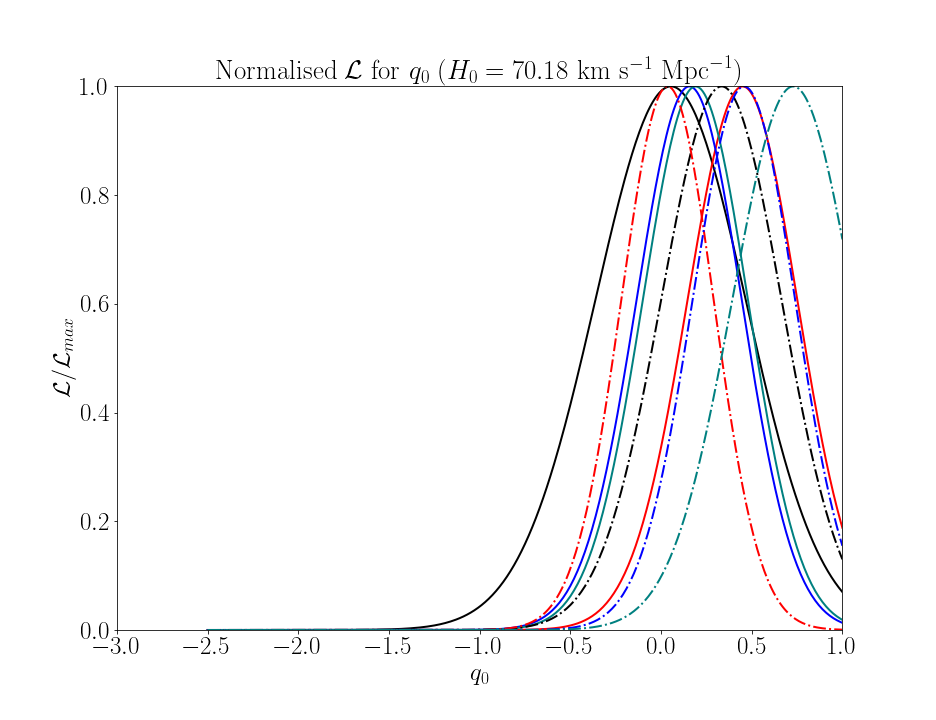}
\includegraphics[width=0.48\linewidth, height=6.0cm]{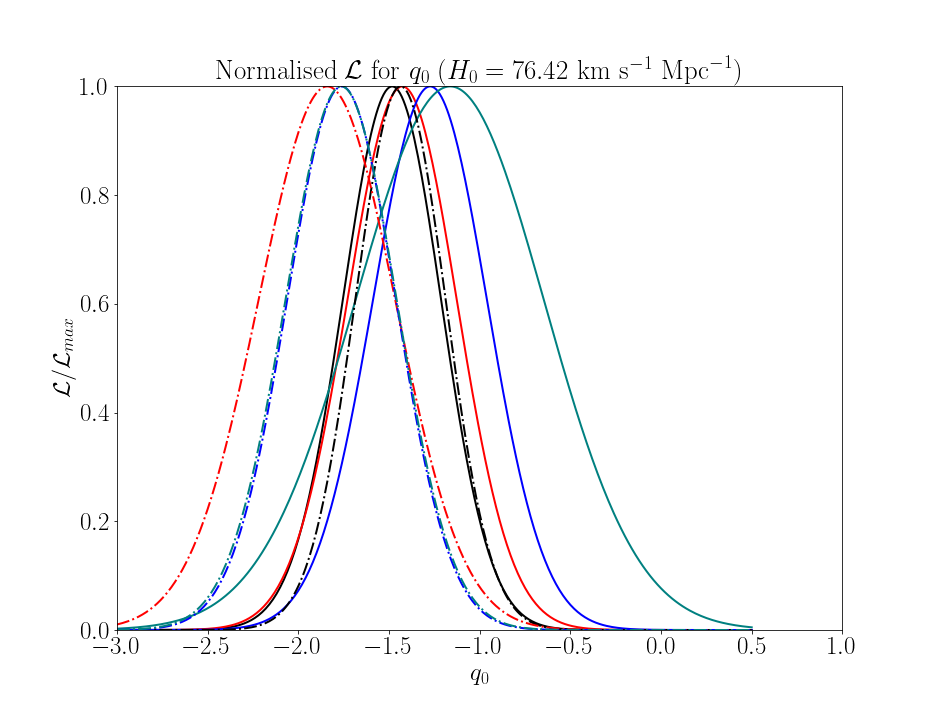}
\caption{Same as left panel of Fig.~\ref{fig:q0lik_deltaq0}, but rather for the normalised likelihoods for $q_0$ when assuming different Hubble Constant values, i.e., $H_0=70.18$ $\mathrm{km}$ $\mathrm{s}^{-1}$ $\mathrm{Mpc}^{-1}$ (left panel), and $H_0=76.42$ $\mathrm{km}$ $\mathrm{s}^{-1}$ $\mathrm{Mpc}^{-1}$ (right panel).}
\label{fig:q0lik_diffH0}
\end{figure*}

As for the statistical significance of our results, we test three main cases: {\bf (i)} How the $\Delta_{q0}$ values can be affected under the assumption of different $H_0$ values to be fixed in our analysis; {\bf (ii)} How the $\Delta_{q0}$ values can be affected when assuming a fiducial $q_0$ value; {\bf (iii)} How the $\Delta_{q0}$ values can be affected if the SN celestial distribution is considered uniform. In all three cases, we stress that we are assuming the same SN covariance matrix as the original one, apart from the redshift selection imposed. 

\begin{figure*}[!ht]
\centering
\includegraphics[width=0.48\linewidth, height=6.3cm]{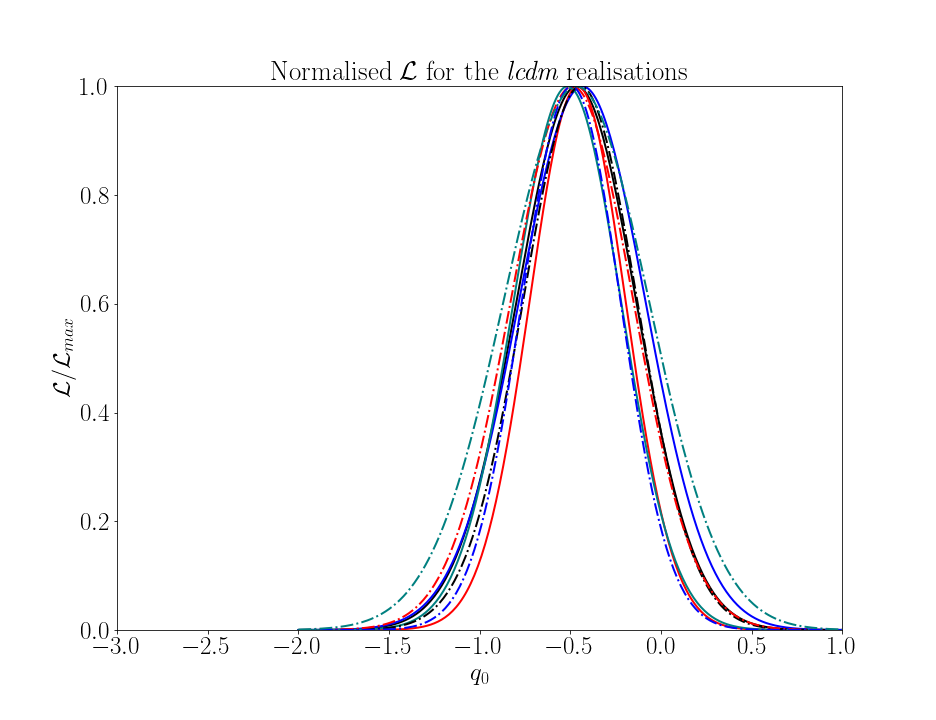}
\includegraphics[width=0.48\linewidth, height=6.3cm]{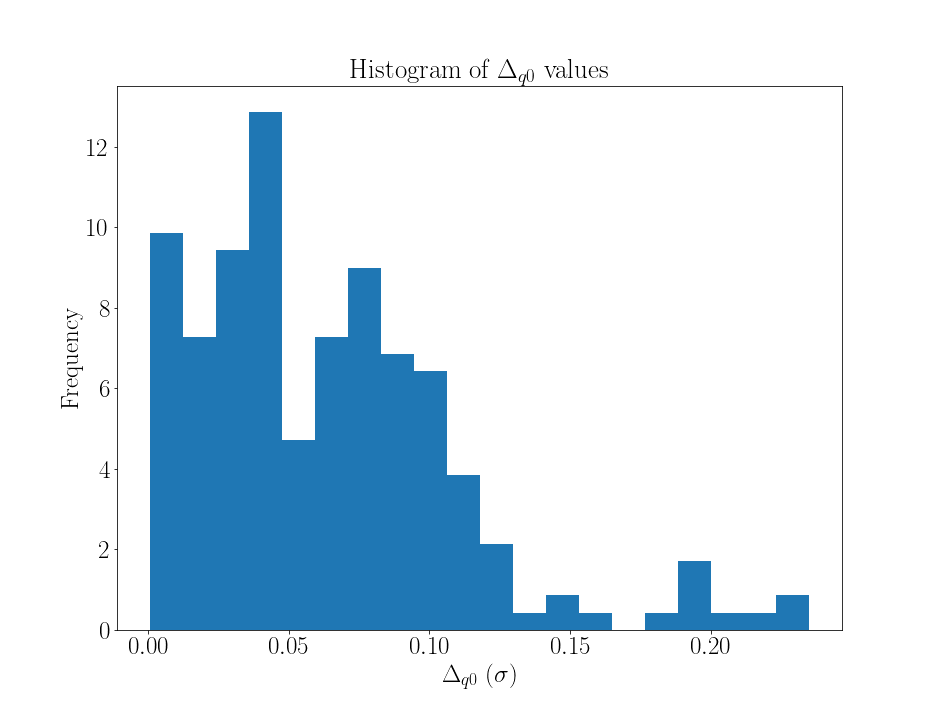}
\caption{Same as Fig.~\ref{fig:q0lik_deltaq0}, but rather for the {\it lcdm} realisations.}
\label{fig:q0lik_deltaq0_lcdm}
\end{figure*}

\begin{figure*}[!ht]
\centering
\includegraphics[width=0.48\linewidth, height=6.3cm]{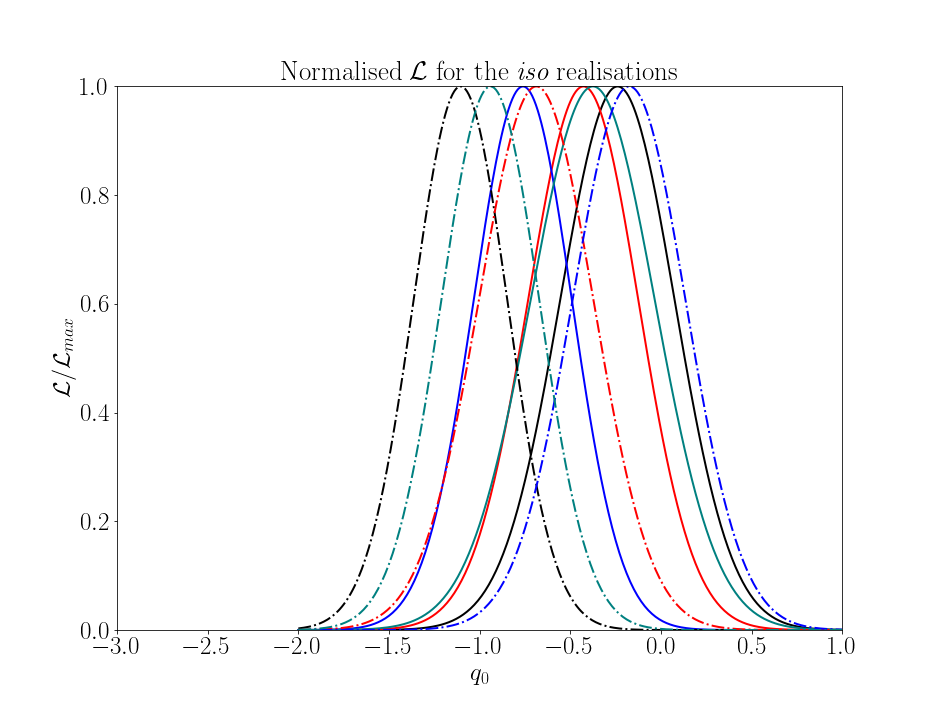}
\includegraphics[width=0.48\linewidth, height=6.3cm]{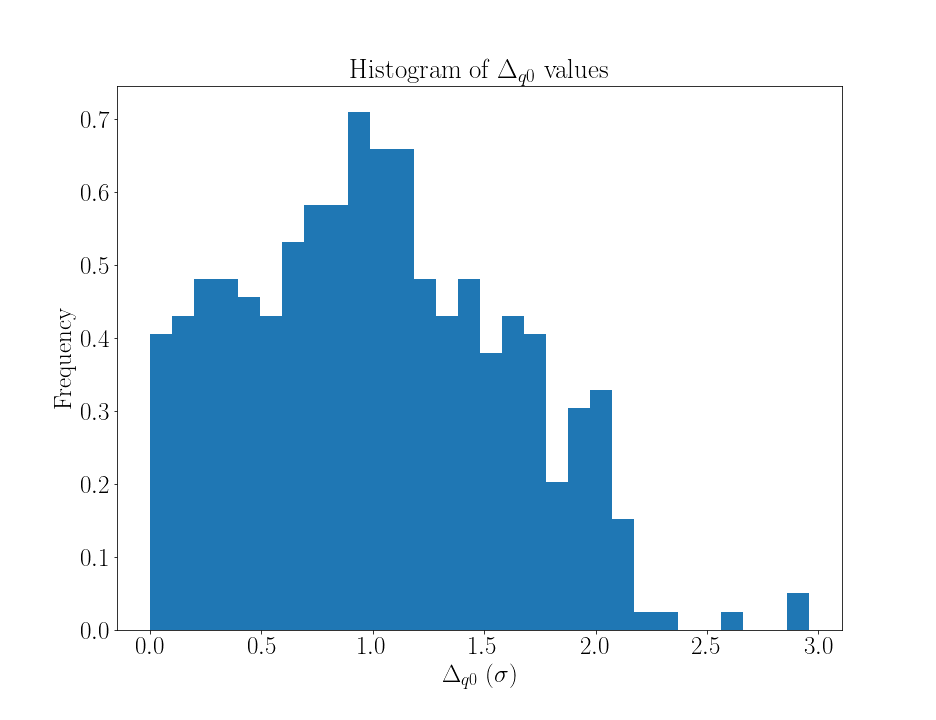}
\caption{Same as Fig.~\ref{fig:q0lik_deltaq0_lcdm}, but rather for the {\it iso} realisations.}
\label{fig:q0lik_deltaq0_iso}
\end{figure*}

As for case {\bf (i)}: In Fig.~\ref{fig:q0lik_diffH0}, we display the normalised likelihoods for $q_0$ when assuming different Hubble Constant values, i.e., $H_0=70.18$ $\mathrm{km}$ $\mathrm{s}^{-1}$ $\mathrm{Mpc}^{-1}$ (left panel), and $H_0=76.42$ $\mathrm{km}$ $\mathrm{s}^{-1}$ $\mathrm{Mpc}^{-1}$ (right panel). We find that a higher $H_0$ value leads to highly negative values for the deceleration parameter. In contrast, the opposite trend occurs for lower $H_0$ values – see how those likelihoods are skewed to the right and left, respectively, compared to those shown in the left panel of Fig.~\ref{fig:q0lik_deltaq0}, where we assumed the SH0ES $H_0$ measurement. Hence, we find that the $q_0$ fit is very sensitive to the $H_0$ assumption, which is an expected result, since this parameter is degenerated with the absolute magnitude $M_B$ value -- and so we would need to adjust $M_B$ accordingly to the change in the $H_0$. We will leave a more thorough examination of the possible degeneracy in the $(M_B,H_0,q_0)$ plane in a future work. 

As for case {\bf (ii)}: In Fig.~\ref{fig:q0lik_deltaq0_lcdm}, we show the results obtained by means of a set of 200 realisations named {\it lcdm}. In this case, we fixed the observed modulus distance to the value given by the flat $\Lambda$CDM best-fit for $q_0$ from the original Pantheon+SH0ES, i.e., $q_0=-0.499$, as $\Omega_{m0}=0.334$ and thus $q_0 = (3/2)\Omega_{m0}-1 = -0.499$. The goal is to assess the residuals of our $q_0$ best-fit estimator due to the limited sampling of SNe across the sky given our $60^{\circ}$ patch size. As we can see in its left panel, we are able to robustly recover the fiducial $q_0$ best-fit in all cases, albeit with larger uncertainty in some of them -- which naturally occur due to the SN sky sampling in those specific patches. On the other hand, the histogram displayed in the right panel shows a maximum value of $\Delta_{q0}=0.24$ in those realisations. We interpret this result as the maximum variance we can expect from our analysis due only to the effect of the SN sample incompleteness across the celestial sphere. 

As for case {\bf (iii)}: In Fig.~\ref{fig:q0lik_deltaq0_iso}, we present normalised $q_0$ for 5 realisations (left panel), and the $\Delta_{q0}$ histogram (right panel) for a set of 400 {\it iso} realisations. Conversely, from the previous cases,
we assume a uniform SN sky distribution by replacing the original SN coordinates with a random direction across the celestial sphere. We obtain a maximum value of $\Delta_{q0}=2.96$ in this case. Such a result is in good agreement with the $\Delta_{q0}=3.06$ result obtained for the real data case -- especially considering the $\Delta_{q0}=0.24$ residual obtained for the {\it lcdm} realisations, as previously described. Therefore, we conclude that there is no statistically significant indication for a breakdown of the cosmic isotropy hypothesis in this case. 

\section{Discussion and conclusions}

The validity of the Cosmological Principle, i.e., the large-scale isotropy and homogeneity of the Universe, constitutes one of the core assumptions of modern Cosmology. Even though the available observational data favours the $\Lambda$CDM scenario based on this hypothesis, it has seldom been tested directly. Hence, developing and performing such tests is crucial since any hint of a statistically significant breakdown of such an assumption would require an extensive reformulation of the standard cosmological model from the basics.

We used the latest Type Ia Supernova compilation of distance measurements, namely the
Pantheon+SH0ES data set to test the isotropy of the Universe. We did so by assessing the directional dependence of the deceleration parameter, where we split the data into sub-sets across randomly selected directions in the sky, in which we obtained their best-fitted values. After restricting our data to the $0.01 < z < 0.10$ range, except for the SN in Cepheid host galaxies, we found a maximum variation of $\Delta_{q0} = 3.06$ at roughly $90^{\circ}$ from the CMB dipole direction – a signal ascribed to our relative motion concerning its rest frame. Moreover, we found that this result does not manifest from the assumptions made during the parameter estimation – e.g., by fixing the Hubble Constant and SN absolute magnitude to its default values or by the incoherent fitting of the deceleration parameter – and most importantly, we found that such a variation is in good agreement with simulations assuming a uniform sky distribution of the SN data points. Therefore, we can conclude that this result is not statistically significant and that it should occur due to intrinsic fluctuations in the data – primarily due to the uncertainties in its covariance matrix.

Our results present an improvement from analyses made on the previous SN compilation, i.e., the Pantheon data set, where it was found from similar tests that the directional dependence of the cosmological parameters could be ascribed to the inhomogeneous SN celestial distribution - see, for instance, ~\cite{Andrade:2018eta}. In addition, we extend and complement former analyses in the literature ~\cite{Pasten:2023rpc,McConville:2023xav,Perivolaropoulos:2023tdt,Tang:2023kzs,Hu:2023eyf}, as we avoided the assumption of dark energy using the cosmographic expansion\footnote{We note that future SN distance measurements might have to assume a higher order cosmographic expansion, as their respective uncertainties might get smaller than the $0.35$\% difference between the predictions from the cosmography expansion up to the second order in redshift, as assumed here, and the $\Lambda$CDM predictions.}, and performed more stringent cuts in the sample to avoid possible biases. Hence, in contrast with some of those results, we found no significant evidence for a possible deviation from the Cosmological Principle assumption in the Pantheon+SH0ES data. This is in good agreement with previous tests using other types cosmological observations, e.g. galaxy clusters~\cite{Bengaly:2015xkw}, infrared galaxies~\cite{Bengaly:2017zlo}, Gamma Ray Bursts~\cite{Andrade:2019kvl}, and quasars~\cite{Mittal:2023xub}. 

Given the advent of ongoing and forthcoming redshift surveys, such as eROSITA~\cite{Merloni:2024zgn}, Vera C. Rubin Observatory~\cite{LSSTScience:2009jmu}, Euclid~\cite{Amendola:2016saw}, Square Kilometer Array Observatory~\cite{SKA:2018ckk}, besides distance measurements by standard sirens from LIGO and Einstein Telescope, we expect that this variance should be reduced even further, and thus, we will be able to pinpoint if the Cosmological Principle provides a realistic representation of the Universe at large scales.


{\it Acknowledgments:} 
CB acknowledges financial support from Funda\c{c}\~ao \`a Pesquisa do Estado do Rio de Janeiro (FAPERJ) - Postdoc Recente Nota 10 (PDR10) fellowship. JSA is supported by CNPq grant No. 307683/2022-2 and Funda\c{c}\~ao de Amparo \`a Pesquisa do Estado do Rio de Janeiro (FAPERJ) grant No. 259610 (2021). This work was developed thanks to the National Observatory Data Center (CPDON).


\end{document}